\documentclass[12pt]{article}

\usepackage{amsmath}
\usepackage{amssymb}
\usepackage{pict2e}        

\usepackage{float}
\usepackage{framed}
\usepackage{floatpag}     

\usepackage{hyperref}
\usepackage{breakurl}
\usepackage{amsmath}
\usepackage{amsfonts}
\usepackage{amssymb}
\usepackage{algorithm}
\usepackage{color}
\usepackage{ntheorem}

\usepackage{hyperref}

\usepackage{pict2e}        
\usepackage[pdftex]{graphicx}
\DeclareGraphicsExtensions{.pdf,.jpeg,.png}

\newcommand{\comment}[1]{}

\def\denseformat{
\setlength{\textheight}{9.2in}
\setlength{\textwidth}{7.1in}
\setlength{\evensidemargin}{-0.2in}
\setlength{\oddsidemargin}{-0.2in}
\setlength{\headsep}{10pt}
\setlength{\topmargin}{-0.3in}
\setlength{\columnsep}{0.375in}
\setlength{\itemsep}{0pt}
\renewcommand{\baselinestretch}{0.99}
}
\def\midformat{
\setlength{\textheight}{8.9in}
\setlength{\textwidth}{6.7in}
\setlength{\evensidemargin}{-0.19in}
\setlength{\oddsidemargin}{-0.19in}
\setlength{\headheight}{0in}
\setlength{\headsep}{10pt}
\setlength{\topsep}{0in}
\setlength{\topmargin}{0.0in}
\setlength{\itemsep}{0in}       
\renewcommand{\baselinestretch}{1.1}
\parskip=0.070in
}
\def\spacyformat{
\setlength{\textheight}{8.8in}
\setlength{\textwidth}{6.5in}
\setlength{\evensidemargin}{-0.18in}
\setlength{\oddsidemargin}{-0.18in}
\setlength{\headheight}{0in}
\setlength{\headsep}{10pt}
\setlength{\topsep}{0in}
\setlength{\topmargin}{0.0in}
\setlength{\itemsep}{0in}      
\renewcommand{\baselinestretch}{1.2}
\parskip=0.080in
}

\def\thisformat{
\setlength{\textheight}{8.8in}
\setlength{\textwidth}{5.5in}
\setlength{\evensidemargin}{0.32in}
\setlength{\oddsidemargin}{0.32in}
\setlength{\headheight}{0in}
\setlength{\headsep}{10pt}
\setlength{\topsep}{0in}
\setlength{\topmargin}{0.0in}
\setlength{\itemsep}{0in}      
\renewcommand{\baselinestretch}{1.2}
\parskip=0.080in
}

\midformat
\spacyformat
\thisformat
\denseformat

\newcommand{\commentOut}[1]{}
\newcommand{\beq}{\begin{equation}}
\newcommand{\eeq}{\end{equation}}
\def\EE{\hbox{I\kern-.1667em\hbox{E}}}

\begin{document}

\bibliographystyle{plain} 

  \title{Consistent Sampling with Replacement}
  \author{Ronald L. Rivest\\
    MIT CSAIL \\
    {\url{rivest@mit.edu}}
  }

  \date{\today}

\maketitle

\begin{abstract}
  We describe a very simple method for ``consistent sampling'' that allows
  for sampling with replacement.
  The method extends previous approaches to consistent sampling, which
  assign a pseudorandom real number to each element, and sample those with
  the smallest associated numbers.  When sampling with replacement, our
  extension gives the item sampled a new, larger, associated
  pseudorandom number, and returns it to the pool of items being sampled.

\end{abstract}  

\section{Introduction}
\label{sec:introduction}

We describe a simple method for ``consistent sampling'' that extends
previous methods to handle sampling with replacement.  We describe
the method and an open-source implementation.

\paragraph{Notation}

Let $\mathcal{I}$ denote a finite nonempty population $\mathcal{I} =
\{1, 2, \ldots, n\}$ of $n$ items from which we wish to draw a sample
\[
    S = S(\mathcal{I}, u, s)
\]
of size $s$, where $u$ is a seed drawn at random from some
large universe $\mathcal{U}$ of seeds.  We emphasize that
$u$ is the only source of randomness for the sampling procedure;
once $u$ is specified the sampling process is deterministic.

Sampling may be performed ``with replacement'' or ``without
replacement.''  When desired to distinguish these cases we
give a superscript ``$+$'' or ``$-$'' to indicate sampling
with or without replacement, as in
\[
    S^+(\mathcal{I}, u, s)
\]
or
\[
    S^-(\mathcal{I}, u, s)\ .
\]

If no superscript is given, the sample may be either with
replacement or without replacement.

When sampling is done with replacement the result is a
\emph{multiset} (set with multiplicities).

The sampling method should be \emph{random} in the sense
that the result of picking seed $u$ at random and then
drawing the sample
\[
    S(\mathcal{I}, u, s)
\]
should result in a \emph{simple random sample} (possibly \emph{with
  replacement})
of size $s$ of $\mathcal{I}$.   (The literature generally uses the term
\emph{simple random sample} to refer to the case where sampling
is done without replacement; the term \emph{simple random sample
  with replacement} is then used to clarify when sampling is done with
  replacement.)

\subsection{Consistent Sampling}
\label{sec:consistent-sampling}

We say that a sampling method $S$ is ``consistent'' if it is
consistent in two ways:
\begin{itemize}

\item It is ``\emph{consistent with respect to sample size}''.
  That is, for any $\mathcal{I}$ and any $u$, we have that
  for any $s$ and $s'$ with $s'\ge s$:
  \[
      S(\mathcal{I}, u, s) \subseteq S(\mathcal{I}, u, s')\ ,
  \]
  so that a larger sample is just an extension of a smaller sample.

  That is, consistency with respect to sample size implies that the
  sampling routine draws elements one at a time from $\mathcal{I}$
  in a particular sequence depending on the seed $u$;
  the sampling is finished when a total of $s$ elements have been drawn.

  With a slight overload of notation, we let $S(\mathcal{I}, u)$
  denote the full sequence of outputs produced by $S$ for a given
  seed $u$: these are the
  elements produced by $S$ as $s$ increases, for $s=1, 2, \ldots$\ .

  If we are sampling without replacement then $S^-(\mathcal{I}, u)$ is
  a finite sequence of length $n=|\mathcal{I}|$.
  If we are sampling with replacement, then $S^+(\mathcal{I}, u)$ is
  an infinite sequence.

\item It is ``\emph{consistent with respect to population}''.
  That is, for any two nonempty sets $\mathcal{J}$ and $\mathcal{K}$
  with $\mathcal{J}\subseteq \mathcal{K}$, we have
  \[
      S(\mathcal{J}, u) = S(\mathcal{K}, u) \cap \mathcal{J}
  \]
  where
  \[
        S \cap \mathcal{J}
  \]
  denotes the subsequence of sequence $S$ obtained by retaining only
  elements in $\mathcal{J}$.
\end{itemize}

\subsection{Proposed Method}

We associate with each item $i$ with a pseudorandom ``(first)
ticket number'' $\tau_{i,1} = f(i,u)$, where $u$ is a random seed.
These ticket numbers are uniformly and independently distributed in
the real interval $(0,1)$, for any fixed $i$, as $u$ varies.

To draw a sample from $\mathcal{I}$ without replacement,
we draw them in order of increasing ticket number.

See
Wikipedia\footnote{\url{https://en.wikipedia.org/wiki/Simple_random_sample}}
for a prior use of this metaphor of ``ticket numbers.''

To sample \emph{with replacement}, when an item $i$ is drawn for the
$j$th time, where $j>1$, it receives a new ticket number $\tau_{i,j} =
g(\tau_{i,j-1})$, where $g$ is a pseudorandom function mapping each real
number $x$ in $(0,1)$
to a value $g(x)$ from the interval $(x,1)$ that (while pseudorandom)
appears to have been drawn at random from $(x,1)$.

We more generally assume that for any $i$ the sequence
\[
    \tau_{i, 1}, \tau_{i, 2}, \tau_{i,3}, \ldots
\]
is indistinguishable from a sequence
\begin{equation}
    x_1, x_2, x_3, \ldots
\label{eqn:xs}
\end{equation}
where $x_1$ is chosen uniformly from the real interval $(0,1)$
and for $j>1$, $x_{j}$ is chosen uniformly from the interval
$(x_{j-1}, 1)$.

See Figure~\ref{fig:main}.

\begin{figure*}
  \floatpagestyle{empty}    
  \framebox[7.1in]{
  \parbox{6.4in}{
\noindent \textbf{\large Consistent Sampling Method}\\[0.3in]
\noindent \textbf{Input:}
          integer $n$,
          random seed $u$,
          integer $s$,
          boolean \textsc{with\_replacement}.
\\[0.2in]
\noindent \textbf{Output:}
A random sample $S$ of size $s$ of $\{1, \ldots, n\}$,
drawn with replacement if input \textsc{with\_replacement} is True.
\\[0.3in]
\noindent \textbf{Method:}
\begin{enumerate}
\item Create a ``first ticket'' $(\tau_{i,1}, i, 1)$ for each item $i$,
  for $i=1, 2, \ldots, n$ where the first ticket number
  \[
  \tau_{i,1} = f(i, u)
  \]
  is the result of applying pseudorandom
  function $f$ to inputs $i$ and $u$ to yield a result uniformly
  distributed in the interval $(0,1)$ (for any fixed $i$, as $u$ varies).
\item Initialize priority-queue (min-heap) $Q$ with the set of tickets so created,
  keyed with their ticket numbers.
\item Initialize the sample $S$ to be the empty set $\phi$.
\item While $S$ has size less than $s$:
  \begin{enumerate}
  \item Extract from $Q$ the ticket $t$ with the least ticket number.
  \item Let $t = (\tau_{i,j}, i, j)$. Place item $i$ into set $S$.
  \item If we are drawing with replacement (that is, if \textsc{with\_replacement} is True), then
    \begin{itemize}
      \item Add ticket $t'$ to $Q$, where $t' = (\tau_{i,j+1}, i, j+1)$,
        where
        \[
        \tau_{i, j+1} = g(\tau_{i,j})
        \]
        for a suitable pseudo-random
        function $g$.
    \end{itemize}
  \end{enumerate}
\item Return $S$ as the desired sample of size $s$.
\end{enumerate}
}}
\caption{The proposed consistent sampling method, 
  based on pseudorandom functions $f$ and $g$.
  The priority queue $Q$ contains exactly one
  ticket $(\tau, i, j)$ for each item $i$.  The value~$j$ is
  the ``generation number'' for the ticket, saying how many
  tickets have been generated for this item so far.  If we are
  sampling without replacement, tickets only have generation
  number~$1$.  Otherwise, tickets with generation number greater
  than~$1$ are replacement tickets.
}
\label{fig:main}
\end{figure*}

The method works for sampling with replacement, since
when a ticket with number $\tau$ is drawn from $Q$ because it
has the minimum ticket number, then all of the remaining tickets
in $Q$ have ticket numbers that are
uniformly distributed in $(\tau, 1)$, conditioned on having just
drawn a ticket with number $\tau$.  So adding a replacement ticket
with ticket number drawn uniformly from $(\tau, 1)$ makes the
new ticket indistinguishable from those already there.

Another useful way of looking what happens with sampling
replacement is to view $Q$ as being initialized with an
infinite number of tickets for each item, one for each
possible generation.  Then sampling from this $Q$ without
replacement is equivalent to sampling from the original~$Q$
with replacement.

Suitable functions $f$ and $g$ are constructible from, say, the
cryptographic hash function SHA256.  (See Section~\ref{sec:implementation}
for details.)  These functions can be implemented in
an efficient manner, with only one or two calls to the underlying
SHA256 hash function required per invocation of $f$ or $g$.
The function $g$ does not need to take seed $u$ as an input if
the ticket numbers are represented in a way that preserves the full
output entropy of the SHA256 hash function.

The consistent sampling method puts the elements of $\mathcal{I}$
into a shuffled order.  A sample of size $s$ is then just the length-$s$
prefix of that order.

The sampling method is consistent.  Note that if $\mathcal{I}$ 
is a population of items, and if $\mathcal{J}$ is a subset of
$\mathcal{I}$, then the order produced for $\mathcal{J}$ is
a subsequence of the order produced for $\mathcal{I}$.

\section{Discussion}
\label{sec:discussion}

The method of assigning a random or pseudorandom number (our ``ticket
number'') to each element is not new, nor is the term ``consistent
sampling.''

The general approach was introduced by
Broder et
al.~\cite{Broder:1997:SCW:283554.283370,broder1997resemblance},
who produced sketches of documents on the web to find
similar documents.  Similarity was estimated by first
computing for each document a sketch consisting of
the set of $s$ features having the smallest hash-value.
Similar documents then have similar sketches.  The
estimates the Jacquard similarity of the two documents.

Recently, Manasse et al.~\cite{Manasse-2010-consistent-weighted}
extended this approach to weighted consistent sampling.

Kutsov et al.~\cite{DBLP:journals/corr/KutzkovP14}
extend consistent sampling to the case where features
are small sets of elements rather than individual elements.

Kane et al.~\cite{Kane-2010-distinct-elements}
apply consistent sampling to the problem of counting the number of
distinct elements in a stream.

Bavarian et
al.~\cite{DBLP:journals/corr/KutzkovP14,DBLP:journals/corr/BavarianGHKRS16}
prove the optimality of such approaches for a certain matching game.

\subsection{Extension to sampling with replacement}

The extension of consistent sampling to handle sampling with
replacement (step 5(c) in Fgure~\ref{fig:main}) appears to be new.

Although our extension (generating a new larger ticket number for an element
when it is sampled with replacement) is very simple and straightforward,
it appears to be irrelevant or unmotivated by previous applications, and
so remained unstudied.

\subsection{Generality}

It is easy to argue, as follows, that the approach taken here is without
loss of generality.

Assume we have some consistent sampling method that works over
subsets of some countable population $\mathcal{I}$.
Let the randomness $u$ be fixed and arbitrary.

Consider the set $V$ of pairs $(i,j)$ where $i\in
\mathcal{I}$ and $j$ is a positive integer.

Given $u$, define a relationship ``$<$'' on $V$ so that
$(i,j)<(i',j')$ if for some $J$ the sampling method on input
$J$ outputs the $j$-th occurence of $i$ at some time before it
outputs the $j'$-th occurence of $i'$.

Consistency implies that (for each fixed $u$) the binary
relation ``$<$'' is a total order on $V$, which
implies (Cantor's Theorem) that $(V,<)$ is isomorphic to a subset of
$Q$ (the rationals).  Thus, we can associate a real number
$\tau_{i,j}$ with each pair $(i,j)$ and output pairs in order of
increasing value $\tau_{i,j}$.  But this is precisely what our
proposed method does.

(To be precise, we have just argued that using ticket numbers
doesn't cause us to miss any opportunities for representing
a consistent sampling method.)

\subsection{Analysis}

It is interesting to ask about the relationship between the number $s$
of items drawn for the sample and the ticket number (call it $\tau_s$)
of the last ticket drawn.  Or similarly, if one draws all items with
ticket number less than a limit $\lambda$, one may be interested in
the distribution of the number $s$ of items drawn.

In this direction, we note that if we define
\begin{equation}
    y_i = 1 - x_i
\end{equation}
where the $x$s are as in (\ref{eqn:xs}), then
$y_k$ is distributed as the product of $k$ independent
uniform variates $z_1$, \ldots, $z_k$.
Since
\[
    \ln(z_i) \sim - \textrm{Exp}(1),
\]
we have
\[
    \ln(y_k) \sim - \textrm{Gamma}(k, 1),
\]
and
\[
    E(\ln(y_k)) = -k\ .
\]
Therefore, if the proposed method is to be used
for sampling with replacement where a given item
may be selected and replaced many (perhaps hundreds)
of times, then the representations of $\tau(i, j)$
should have sufficient precision to handle numbers that
are extremely close to 1 (or if the $y$s are represented
instead of the $x$s, to handle numbers with large negative
exponents).  That is to say, the number of bits needed
to represent $\tau(i,j)$ grows linearly with $j$.

\section{Implementation}
\label{sec:implementation}

Python 3 code for this method is given in Github:
\begin{quote}
  \url{https://github.com/ron-rivest/consistent_sampler}
  \end{quote}

The representation of ticket numbers in this python code
uses variable-length numbers (represented as decimal strings)
with no upper limit on the precision.

We note that for sampling with replacement the implementation
picks a pseudorandom $y$ in the range $(x, 1)$ by:
\begin{enumerate}
\item Obtaining $x'$ by deleting all digits in $x$ after the
  initial segment of 9's.  For example,
  $x=0.99995241$ becomes $x'=0.9999$.
  Set counter $i$ to $1$.
\item Generating a uniform pseudorandom variate $v$ by
  hashing $x$ and $i$.  Then increase $i$ by one.
  Example: $v=0.77318824$.
\item Creating a candidate $y$ by appending the digits of $v$
  to the end of $x'$.  Example: $y=0.999977318824$.
\item Returning $y$ if it is larger than $x$.  Otherwise
  return to step 2 and repeat.
\end{enumerate}
This approach is quite portable, and avoids having to do
high-precision multiplication.  The expected number of iterations
of this loop to obtain a value $y$ that is larger than $x$ depends
on $x$, but is not more than ten, and has expected value
$3.143$.

The efficiency of the method is determined by the efficiency
of SHA256, which is called once to compute each initial
ticket number, and about $3.14$ times for each replacement
ticket number.
A typical laptop can compute about
one million SHA256 hash values per second.

\section*{Acknowledgments}
\label{sec:acknowledgments}

The author gratefully acknowledge support for their work on this
project received from the Center for Science of Information (CSoI), an
NSF Science and Technology Center, under grant agreement CCF-0939370.

\bibliography{cs,references}

\end{document}